\documentclass[a4paper,fleqn,usenatbib]{mnras}     

\usepackage{ae,aecompl}

\usepackage{graphicx}	
\usepackage{amsmath}	
\usepackage{amssymb}	
\usepackage{pdflscape}
\usepackage{float}

\def\braket#1{\left<#1\right>}
\def\Dnu{\Delta{\nu}}
\def\bDnu{\braket{\Delta{\nu}}}

\title[solar-like oscillating evolved stars]{Asteroseismic analysis of 15 solar-like oscillating evolved stars}
\author[Z. \c{C}elik Orhan, M. Y{\i}ld{\i}z and C. Kayhan ]{Z. \c{C}elik Orhan,$^{1}$\thanks{Contact e-mail: \href{mailto:mn@ras.org.uk}{zeynep.celik@ege.edu.tr}} M. Y{\i}ld{\i}z $^{1}$ and C.  Kayhan$^{1,2}$
\\
$^{1}$Department of Astronomy and Space Sciences, Science Faculty, Ege University, 35100, Bornova, \.{I}zmir, Turkey\\
$^{2}$Department of Astronomy and Space Sciences, Science Faculty, Erciyes University, 38030, Melikgazi, Kayseri, Turkey\\
}
\date{Accepted XXX. Received YYY; in original form ZZZ}

\begin{document}
\label{firstpage}
\pagerange{\pageref{firstpage}--\pageref{lastpage}}
\maketitle

\begin{abstract}
Asteroseismology using space-based telescopes is vital to our understanding of stellar structure and evolution. {\textit{CoRoT}}, {\textit{Kepler}}, and {\textit{TESS}} space telescopes have detected large numbers of solar-like oscillating evolved stars. 
Solar-like oscillation frequencies have an important role in the determination of fundamental stellar parameters; in the literature, the relations between the two is established by the so-called scaling relations. 
In this study, we analyse data obtained from the observation of 15 evolved solar-like oscillating stars using the {\textit{Kepler}} and ground-based 
telescopes.
 The main purpose of the study is to determine very precisely the fundamental parameters of evolved stars by constructing interior models using asteroseismic parameters.
We also fit the reference frequencies of models to the  observational reference frequencies caused by the He {\scriptsize II} ionization zone. 
 The 15 evolved stars are found to have masses and radii within ranges of $0.79$-$1.47$ $M_{\rm \sun}$ and $1.60$-$3.15$ $R_{\rm \sun}$, respectively.
Their model ages range from $2.19$ to $12.75$ Gyr.
It is revealed that fitting reference frequencies typically increase the accuracy of asteroseismic radius, mass, and age.
The typical uncertainties of mass and radius are $\sim$ 3-6 and $\sim$ 1-2 per cent, respectively.
Accordingly, the differences between  the model and literature ages are generally only a few Gyr. 

\end{abstract}




\section{Introduction}

The determination of solar-like oscillation frequencies using information gathered by space telescopes is crucial
to our understanding of stellar evolution and structure. 
Such oscillations are excited in stars with convective envelopes and 
have been detected in low-mass main-sequence (MS), 
subgiant (SG), and red giant (RG) stars.  
Connecting interior stellar models to measured oscillation frequencies as well as other measurements, such as those obtained from spectroscopy, 
yields precise determinations of stellar parameters such as age ($t$), mass ($M$), and radius ($R$).
It was also confirmed that fitting model reference frequencies to the observed reference frequencies decreases uncertainty in asteroseismic $M$, $R$, and $t$  
in comparison with uncertainties of the same parameters obtained by fitting large separations between oscillation frequencies ($\Dnu$) and the frequency of maximum amplitude ($\nu _{\rm max}$).
Model stellar $M$ and $R$ are approximately several times more accurate than  the $M$ and $R$ obtained from the scaling relations.
Precisely determined fundamental parameters are broadly 
useful for a variety of efforts in astrophysics to test and develop 
theories of stellar and galactic evolution
(e.g. Deheuvels et al. 2016; Nissen et al. 2017; Bellinger et al. 2017), and to understand the formation and evolution of exo-planetary systems (Chaplin and Miglio 2013,
Campante et al. 2015, Campante et al. 2016, Kayhan, Y\i ld\i z \& \c{C}elik Orhan 2019, Jiang et al. 2020).

In this study, we constructed interior models of 15 evolved targets using the {{\small MESA}} code (Paxton et al.  2011, 2013) and 
used them to determine the stars' fundamental properties under asteroseismic 
and non-asteroseismic constraints. 
The asteroseismic constraints used for calibration of interior models, which are derived from {\textit{Kepler}} light curves for 14 targets (see Table 1 for the references) and from ground-based observations of one target 
(HD 2151,  $\beta$ Hyi, Bedding et al. 2007), included the reference frequencies discovered by Y\i ld\i z et al. (2014a) 
and customary asteroseismic parameters such as the $\Dnu$ and $\nu _{\rm max}$.

In the literature, there are many studies on the determination of the fundamental parameters of solar-like oscillating stars based on 
 fitting of model adiabatic oscillation frequencies to observed frequencies (Li et al. 2020,  Kayhan, Y\i ld\i z \& \c{C}elik Orhan 2019, Metcalfe et al. 2014, Mathur et al. 2012) and on the relating of asteroseismic parameters to non-asteroseismic properties 
(Y\i ld\i z, \c{C}elik Orhan \& Kayhan 2019; hereafter Paper IV, Mathur et al. 2012, White et al. 2011).  
Tassoul (1980), Ulrich (1986) and Brown et al. (1991)
have developed the general scaling relations between global asteroseismic observable parameters 
and fundamental stellar properties; subsequently Kjeldsen $\&$ Bedding (1995) scaled stellar properties to values observed in 
the Sun. We call these relations as the conventional scaling relations. 
$M_{\rm sca}$ and $R_{\rm sca}$ can be given as functions of 
mean $\Delta \nu$ ($\braket{\Delta\nu}$), $\nu_{\rm max}$ and 
effective temperature (${T_{\rm eff}}$) as follows:
\begin{equation}
\frac{M_{\rm sca}}{M_{\rm \odot}}= \left(\frac{\nu_{\rm max}}{\nu_{\rm max\odot}}\right)^3\left(\frac{{\braket{\Delta\nu_{\rm }}}}{{\braket{\Delta\nu_{\rm \odot}}}}\right)^{-4}\left(\frac{T_{\rm eff}}{T_{\rm eff \odot}}\right)^{1.5} 
\end{equation}
and
\begin{equation}
\frac{R_{\rm sca}}{R_{\rm \odot}}= \left(\frac{\nu_{\rm max}}{\nu_{\rm max\odot}}\right)\left(\frac{{\braket{\Delta\nu_{\rm }}}}{{\braket{\Delta\nu_{\rm \odot}}}}\right)^{-2}\left(\frac{T_{\rm eff}}{T_{\rm eff \odot}}\right)^{0.5}, 
\end{equation}
where ${\nu_{\rm max, \odot}}$ and $\braket{\Delta\nu_{\rm \odot}}$ are  
the solar values of ${\nu_{\rm max}}$  and $\braket{\Delta\nu}$, respectively, 
and are taken as ${\nu_{\rm max, \odot}}$=3050 ${\rm \mu}$Hz (Kjeldsen $\&$ Bedding 1995) and $\braket{\Delta\nu_{\rm \odot}}$=135.15 ${\rm \mu }$Hz (Chaplin et al. 2014). 

\par These conventional scaling relations inherently assume that the Sun and the other solar-like oscillating stars have 
similar internal structures, an assumption that has still been seriously tested for the stars in various evolutionary phases. 
Accordingly, using these assumptions to investigate stars with varying stellar structures and in different evolutionary phases can lead to systematic errors 
in the values of $M$ and $R$ obtained from the scaling relations 
(see e.g. Y{\i}ld{\i}z et al. 2015; hereafter Paper II, Epstein et al. 2014, Chaplin \& 
Miglio 2013, Corsaro et al. 2013, Huber et al. 2013, Miglio et al. 2012).
In the literature, many approaches have been proposed to reduce 
these systematic errors ( Zinn et al. 2019, Sahlholdt et al. 2018, Viani et al. 2017, Huber et al. 2017, Sharma et al. 2016, Y{\i}ld{\i}z, \c{C}elik Orhan \& Kayhan 2016; hereafter Paper III, Guggenberger et al. 2016, Y{\i}ld{\i}z et al. 2014a; hereafter Paper I, Mosser et al. 2013, White et al. 2011).

Because they have outer and inner structures that differ significantly from those in the Sun, it is particularly important to test the scaling relations for evolved stars.
By calibrating these models, we are able to determine the $M$, $R$, luminosities ($L$), and $t$ of the targets with high precision.
To this end, we select and analyse fourteen solar-like evolved stars detected by \textit{Kepler} and  one star detected though ground-based observation (HD 2151) and   
used the obtained data to calculate their values of 
$M$ and $R$ by applying equations (1) and (2), 
respectively. We then constructed interior models of the stars  using the {\small {MESA}} evolution code, 
to reobtain $M$, $R$, and the other stellar parameters. 

Oscillation frequencies of the {\small {MESA}} models are computed using the {\small {ADIPLS}} pulsation package 
(Christensen-Dalsgaard, 2008). 
We determine reference frequencies 
(${\nu_{\rm min0}}$, ${\nu_{\rm min1}}$, ${\nu_{\rm min2}}$) using the methods given in Paper I and Paper II, and  
$\braket{\Delta\nu}$ 
from observed and model oscillation frequencies. 
The typical uncertainties for $M$ and $R$ are found to be approximately 3 and 1 per cent, respectively.

This paper is organized as follows. 
The observational properties of solar-like oscillating evolved stars are presented in Section 2. 
We explain the basic properties of the {\small {MESA}} code and the models used in the study in Section 3. 
In Section 4, we present and discuss the model results, and we compare our fundamental stellar model parameters with those found in the literature. Finally, we present conclusions in Section 5.

\section{PROPERTIES OF SOLAR-LIKE OSCILLATING EVOLVED STARS}
We analyse the solar-like oscillations of the 15 selected evolved stars. 
In Fig. 1, the stars are plotted on the Hertzsprung-Russell (HR) diagram. 
The luminosity values ($L_{\rm sca}$) on the diagram are obtained from asteroseismic values of $R_{\rm sca}$ derived from equation (2) and from the respective spectral effective temperatures ($T_{\rm es}$): 
$L_{\rm sca}=4 \pi \sigma R_{\rm sca}^2 T_{\rm es}^4$, where $\sigma$ is Stefan-Boltzmann constant.
The thin and thick lines represent, respectively, the  zero 
age main-sequence (ZAMS) and  terminal age main-sequence (TAMS) obtained from
{\small ANK\.{I} } models with  solar composition   (Y{\i}ld{\i}z 2015), while
the filled circles indicate the 15  evolved stars.
The observed oscillation frequencies of the targets,  
which included 13 SGs
and 2 RG stars (KIC 7341231 and KIC 8561221),
are obtained with a high degree of precision.
Asteroseismic and non-asteroseismic observational data for the stars are listed in Table 1.

\begin{figure}
\begin{center}
\includegraphics[width=\columnwidth]{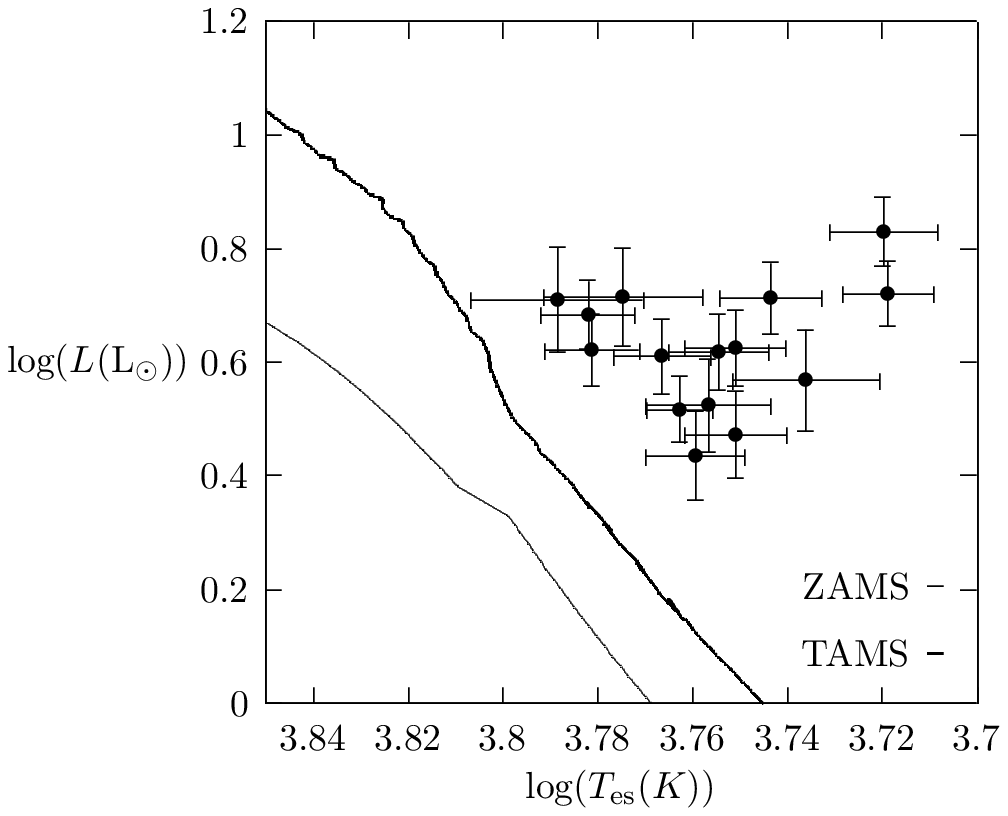}

   \caption{HR diagram for stars observed using \textit{Kepler} and ground-based telescope. 
The thin and thick lines show, respectively, ZAMS and TAMS lines (Y{\i}ld{\i}z 2015); filled circles represent evolved stars analysed in this study.  
 }
\end{center}
\end{figure}

The spectroscopic observational data of the \textit{Kepler} target stars, i.e., the gravity ($\log g$), metallicity ([Fe/H]) and $T_{\rm es}$,  
are  adapted from Bruntt et al. (2012) and Molenda-{\.{Z}}akowicz et al. (2013), while their $\braket{\Delta\nu}$ and $\nu_{\rm \max}$  values are taken from Chaplin et al. (2014).
 In this study,  
observed $T_{\rm es}$ and $\log g$ values selected from different spectral studies in Table 1 are most compatible with $T_{\rm eff}$ and $\log g$ values obtained by different methods in Paper IV.
Thus, when these values obtained from different methods are compared, they are very close to each other (see Fig. 7).
The observed oscillation frequencies of the stars, with the exception of HD 2151 (Bedding et al. 2007), are taken from Appourchaux et al. (2012).

For most stars, oscillation frequencies ($\nu_{nl}$) of many modes with low degrees ($l$) and
high order ($n$) are available, allowing the computation of $\Delta\nu$
and the small separation between oscillation frequencies ($\delta \nu_{02}=\nu_{n0}-\nu_{n-1,2}$).
${\braket{{\delta{\nu_{02}}}}}$ is the mean value of $\delta \nu_{02}$, and listed 
in Table 1.

Determination of the reference frequencies is an efficient method for yielding the mass, radius, and effective temperature of solar-like oscillating stars (Papers I-IV). 
In this study, we obtained observed reference frequencies from a graph of $\Delta\nu$ as a function of $\nu$. 
Although values of $\nu_{\rm min1}$ are available for 14 of the stars,  value of $\nu_{\rm min0}$ is obtained for 12 stars (see Table 1), 
$\nu_{\rm min2}$ values are not obtained for any of the stars.

The reference frequencies are due to glitches caused by the He {\scriptsize II} ionization zone.
To determine these frequencies, we follow the method described in Paper I: after determining the  
frequency interval corresponding to each minimum on the $\Dnu$ versus $\nu$ graph, we draw two lines from the neighboring intervals and
set their intersection as the reference frequency 
$\nu_{\rm min}$  
for that minimum.
 The maximum uncertainties in $\nu_{\rm min0}$ and $\nu_{\rm min1}$ are about half of the uncertainties in large separation (Paper III).
 The uncertainties in $\nu_{\rm min0}$ and $\nu_{\rm min1}$  are given in Table 1.

11  of the 15 target stars- 
HD 2151, KIC 5955122, KIC 7341231, KIC 7747079, KIC 7976303, KIC 8561221, KIC 8702606, KIC 10920273, 
KIC 11026764, KIC 11395018 and KIC 11414712 show mixed modes
in their oscillation frequencies.
Although the remaining stars are evolved, they featured no mixed mode behavior.

The effective temperature range of the stars is 5233 K (KIC 7341231) to 6145 K (KIC 10018963). The [Fe/H] values range from -1.64 (KIC 7341231) to 0.35 dex (KIC 11244118). 

\begin{table*}
\begin{center}
\caption{ Observed properties of evolved solar-like oscillating stars. The listed asteroseismic and non-asteroseismic properties are $T_{\rm es}$, logarithm of surface gravity (cgs), [Fe/H], $\braket{\Delta\nu}$, mean small separation ($\braket{\delta\nu_{\rm 02}}$), ${\nu_{\rm max}}$, ${\nu_{\rm min0}}$, ${\nu_{\rm min1}}$, and references for the  observational data.}
\label{tab:observation}
\centering
\begin{tabular}{lccccccccccccccr}
             \hline
              Star  &$T_{\rm es}$&$\log g$& [Fe/H] &$\braket{\Delta\nu}$ & $\braket{\delta\nu_{\rm 02}}$ &${\nu_{\rm max}}$ & ${\nu_{\rm min0}}$&${\nu_{\rm min1}}$& ref \\
              &(K)&(cgs)&(dex)&(${\rm \mu}$Hz)&(${\rm \mu}$Hz)&(${\rm \mu}$Hz)&(${\rm \mu}$Hz)&(${\rm \mu}$Hz)\\
  \hline
   \verb'HD 2151'  &5873$\pm$45 &3.98$\pm$0.02&-0.11$\pm$0.08&57.5$\pm$0.6 &5.11$\pm$1.21&1000.6$\pm$30&1120.44$\pm$0.30&890.33$\pm$0.30&2,3,4\\[2pt]
   \verb'KIC 5955122'  &5865$\pm$70 &3.88$\pm$0.08&-0.06$\pm$0.06&49.2$\pm$0.9&5.20$\pm$0.31&861.0$\pm$24&952.67$\pm$0.45&721.07$\pm$0.45&1,5,7,13,15\\[2pt]  
  \verb'KIC 7341231'&5233$\pm$50 &3.55$\pm$0.03&-1.64$\pm$0.05&29.2$\pm$0.7&3.40$\pm$0.17&408.0$\pm$8&-----&342.04$\pm$0.35&1,7,9\\[2pt]
   \verb'KIC 7747078' &5840$\pm$70&3.98$\pm$0.08&-0.11$\pm$0.06&53.9$\pm$0.3 &4.67$\pm$0.17&936.0$\pm$32&1037.89$\pm$0.15&793.81$\pm$0.15&1,5,7,14,15\\[2pt]
   \verb'KIC 7976303' &6053$\pm$70 &3.87$\pm$0.08&-0.41$\pm$0.06&51.7$\pm$0.6 &4.46$\pm$0.39&851.0$\pm$20&980.4$\pm$0.30&777.44$\pm$0.30&5,7,13,15\\[2pt]
   \verb'KIC 8228742' &6042$\pm$70 &4.02$\pm$0.08&0.00$\pm$0.06&61.8$\pm$0.6 & 4.80$\pm$0.37&1171.0$\pm$34&------&1038.97$\pm$0.30&1,5,7,13,15\\[2pt]  
  \verb'KIC 8524425' &5634$\pm$70 &3.98$\pm$0.08&0.13$\pm$0.05&59.4$\pm$0.6 &4.95$\pm$0.13& 1081.0$\pm$28& 1128.00$\pm$0.30&897.61$\pm$0.30&1,5,7,14,15\\[2pt]   
   \verb'KIC 8561221' &5245$\pm$70 &3.61$\pm$0.08&-0.04$\pm$0.06&29.6$\pm$0.4 & 2.40$\pm$0.14& 467.0$\pm$23&530.46$\pm$0.20&398.87$\pm$0.20&5,12,15\\[2pt]
   \verb'KIC 8702606' &5540$\pm$70 &3.96$\pm$0.08&-0.06$\pm$0.05&39.9$\pm$0.5 &3.59$\pm$0.16&664.0$\pm$16&688.67$\pm$0.25&530.5$\pm$0.25&1,5,10,15\\[2pt]
   \verb'KIC 10018963' &6145$\pm$112 &3.95$\pm$0.21&-0.16$\pm$0.05&55.5$\pm$0.5 & 5.06$\pm$0.23& 987.0$\pm$32&997.0$\pm$0.25&------&1,7,14,15\\[2pt]
   \verb'KIC 10920273' &5710$\pm$75 &4.15$\pm$0.08&-0.02$\pm$0.07&57.3$\pm$0.1& 4.84$\pm$0.23&990.0$\pm$60&1111.50$\pm$0.05&882.77$\pm$0.05&6,8,11\\[2pt]
  \verb'KIC 11026764' &5682$\pm$70 & 3.88$\pm$ 0.08& 0.11$\pm$0.06&50.5$\pm$0.6 & 4.50$\pm$0.18& 895.0$\pm$29&924.01$\pm$0.30&723.52$\pm$0.30&1,5,7,15\\[2pt] 
   \verb'KIC 11244118' &5745$\pm$70 &4.09$\pm$0.08&0.35$\pm$ 0.06&71.3$\pm$0.9 &5.50$\pm$0.15&1420.0$\pm$31&------&1166.25$\pm$0.45&1,5,7,14,15\\[2pt]
   \verb'KIC 11395018' &5445$\pm$85 &3.84$\pm$0.12&0.13$\pm$0.07&47.5$\pm$0.1& 5.50$\pm$0.30& 834.0$\pm$50& 875.30$\pm$0.05&685.40$\pm$0.05&1,8,11\\[2pt]
   \verb'KIC 11414712' &5635$\pm$70 &3.80$\pm$0.08&-0.02$\pm$0.05&43.5$\pm$0.7 & 4.19$\pm$0.21& 707.0$\pm$20&800.43$\pm$0.35&628.85$\pm$0.35&1,5,7,15\\[2pt] 

  \hline
\end{tabular}
\begin{flushleft}
\textbf{1}: Appourchaux et al.(2012), \textbf{2}: Bedding et al.(2007), \textbf{3}: Brand{\~{a}}o et al.(2011), \textbf{4}: Bruntt et al.(2010), \textbf{5}: Bruntt et al.(2012), \textbf{6}: Campante et al.(2011), \textbf{7}: Chaplin et al.(2014), \textbf{8}: Creevey et al.(2012), \textbf{9}: Deheuvels et al.(2012), \textbf{10}: Deheuvels et al.(2014), \textbf{11}: Do{\u{g}}an et al.(2013),\textbf{12}: Garc{\'{i}}a et al.(2014), \textbf{13}: Mathur et al.(2012), \textbf{14}: Metcalfe et al.(2014), \textbf{15}: Molenda-{\.{Z}}akowicz et al.(2013).

\end{flushleft}
\end{center}
\end{table*}

\section{PROPERTIES OF THE {\small {MESA}} CODE AND MODELS}
\subsection{Properties of the {\small {MESA}} code}

Stellar interior models are constructed using the {\small {MESA}} evolution code. Its version is 7184 (Paxton et al. 2011, 2013). 
Convection is treated using standard mixing-length theory (B\"{o}hm - Vitense 1958). Convective overshooting (for an asteroseismic investigation on this subject, see Angelou et al. 2020)
is not included.
{\small{MESA}} incorporates the opacity tables developed by Iglesias 
\& Rogers (1993, 1996) and includes their OPAL opacity tables in the high-temperature region supplemented by the low-temperature tables produced by Ferguson et al. (2005), with fixed metallicity set
as the default option. The pre-main sequence is included in the construction of the stellar interior models. As the oscillation frequencies of the models are significantly influenced by atmospheric conditions,
 we applied the  \texttt{simple$\_$photosphere} option of the {\small {MESA}} code to construct the interior stellar models. 
The models do not 
include microscopic diffusion; nuclear reaction rates are adopted from Angulo et al. (1999) and Caughlan $\&$ Fowler (1988). The oscillation 
frequencies of the interior models are computed using the {\small{ADIPLS}} pulsation package (Christensen-Dalsgaard 2008).

\subsection{Properties of the {\small {MESA}} models}

The observed properties of the targets are used as either input parameters or constraints for the interior models. 
The input parameters of the {\small {MESA}} evolution code are $M$, helium abundance ($Y$), heavy element abundance ($Z$), and  the convective parameter ($\alpha$). Stellar mass- the most important input parameter- is calculated from the 
scaling relations if no model value  is  available from the literature for the star in question. 
For the initial models,
the solar values of $\alpha$ and $Y$ are used. 
To calibrate the interior model, 
we adjusted the values of $M$, $Y$, and $\alpha$ until a fit for the observed asteroseismic and non-asteroseismic constraints is obtained.   
Based on calibration of the solar model, the values of $Y$, $Z$, and $\alpha$ for the Sun are obtained as $0.2792$, $0.0172$, and $2.175$, respectively. 

In general, the $Z$ of a star is computed from its observed [Fe/H]  value under the assumption that the heavy elements all have the same relative abundance as iron in the solar composition. 
However,
iron is generally  not the most abundant stellar element and is not a good indicator for all heavy elements.
Thus, in the absence of detailed information on chemical composition, this issue can be partially overcome by using oxygen- generally the most abundant heavy element in a star- as superior indicator of total metallicity.
For this purpose, 
Y{\i}ld{\i}z et al. (2014b) analysed the oxygen [O/H] and [Fe/H] abundances produced by 
Edvardsson et al. (1993, see their fig. 15a-1) to derive a relationship between the two using data
within the  [Fe/H]=-0.5-0.5 dex.  
Following Y{\i}ld{\i}z et al. (2014b), we first calculate [O/H] abundances from [Fe/H]
using this relationship and then used the results to obtain the total metallicities ($Z_{\rm O}$).
The model
 metallicities ($Z_{\rm mod}$) of the 15 stars, taken as $Z_{\rm mod}=Z_{\rm O}$, are listed in Table 2.
KIC 7341231, the [Fe/H] is found very low (-1.64) and, therefore, its  total metallicity is set to $0.001$.

The spectroscopic effective temperature and metallicity,
$\bDnu$ and
the reference frequencies 
(and partly small separation between oscillation frequencies, $\braket{\delta \nu_{02}}$, see below) 
are all used as constraints to the interior
models. This means that a calibrated model of any star should have the same
value of these observables within the range of uncertainties. The input parameters $M$, $Y$,  and  $\alpha$ are modified
in an  amount found from comparison of model values with observational
constraints during the calibration procedure:\\
i) At the beginning of calibration, we compute $M_{\rm sca}$ and $R_{\rm sca}$ using equations (1) and (2).
We take initial value of $M$ as either $M_{\rm lit}$ or $M_{\rm sca}$.
For $Y$ and  $\alpha$, the initial values are the solar values. 
During the calibration, $R$ and L are taken as $R_{\rm sca}$ and $L_{\rm sca}$, respectively. 
\\
ii) We calibrate model in the HR diagram by changing $\alpha$,  and then compute $\bDnu_{\rm mod}$, 
$\nu_{\rm min0,mod}$ and $\nu_{\rm min1,mod}$.\\
iii) From comparison of model and observed $\bDnu$s, 
we estimate new value of $M$.
It  is computed from the $\bDnu$-$\rho$ relation:
$(\bDnu_{\rm mod}-\bDnu_{\rm obs})/\bDnu_{\rm mod}= (\Delta M$/$M_{\rm mod})/2$, because $R_{\rm mod}$ is already fitted to $R_{\rm sca}$ in the second step.
If $\bDnu_{\rm mod}$ is fitted to $\bDnu_{\rm obs}$, then $M$ is fixed and we compare model and observed reference frequencies.\\
iv)
$\chi^2_{\rm seis}$ is used in order to examine the fit of the model parameters with the observed parameters.
If $\chi^2_{\rm seis}$ is small enough, $\chi^2_{\rm seis}< 1$, model is calibrated.
If $\chi^2_{\rm seis}$ is not small,
we try to fit model reference frequencies to observed reference frequencies by changing $Y$ and go to the second step.
\\
v) If agreement between model and observations is not satisfactory, we slightly modify $T_{\rm eff}$ and then start from the beginning.
This was performed by staying within the observed error range of $T_{\rm eff}$ change. Increase or decrease in $T_{\rm eff}$ depends on the difference between 
model and observed reference frequencies. For KIC 1126764, for example,  the derivatives $\delta T_{\rm eff}/\delta \nu_{\rm min0}$ and 
$\delta T_{\rm eff}/\delta \nu_{\rm min0}$ are obtained as 2.371 and 1.182 from the interior models, respectively.
Using these derivatives, and differences between the model and observational reference frequencies (${\nu_{\rm min0}}-{\nu_{\rm min0,mod}}$ and ${\nu_{\rm min1}}-{\nu_{\rm min1,mod}}$),
we estimate the new $T_{\rm eff}$ to decrease $\chi^2_{\rm seis}$ and restart from the first step.
As long as $|T_{\rm es}-T_{\rm eff}|$ is less than the uncertainty in $T_{\rm es}$, 
$\chi^2_{\rm spec}< 1$.
\\

$\delta\nu_{\rm 02}$ is the difference between the oscillation frequencies of modes  of consecutive orders with $l=0$ and 2. 
 The cavity of the mode with $l=0$ contains the nuclear core, while the
turning points of the modes with $l=2$ are about the outer border of the nuclear core. 
For $\alpha$ Cen A, for example, the turning points of the observed modes with $l=2$ varies between 0.085 and 0.0125 $R$ (see figure 2 in  Y{\i}ld{\i}z 2008).  
Due to nuclear evolution of the core the value of $\delta\nu_{\rm 02}$ also decreases during 
the MS phase, making it a good age indicator for that phase.  
Although in evolved stars the values of $\delta\nu_{\rm 02}$ is not very sensitive to age, it can still be useful in aging some SGs
by fitting the model values to the observed values.

To quantify the differences between the model results and observational data obtained for each star,
we calculated a normalized $\chi^2$ for asteroseismic and non-asteroseismic constraints: 
%
\begin{equation}
{\chi^2_{\rm seis}}= \frac{1}{N_{\rm f}}\sum\limits_{i=1}^n\left(\frac{{Q_{\rm i,obs}}-{Q_{\rm i,mod}}}{\sigma_{\rm i,obs}}\right)^2
\end{equation}
and
\begin{equation}
{\chi^2_{\rm spec}}= \frac{1}{N_{\rm s}}\sum\limits_{i=1}^n\left(\frac{{P_{\rm i,obs}}-{P_{\rm i,mod}}}{\sigma_{\rm i,obs}}\right)^2
\end{equation}
where ${Q_{\rm i,obs}}$ and ${Q_{\rm i,mod}}$ are the observed and model values,  respectively: $\braket{\Delta\nu}$,  ${\nu_{\rm min0}}$, and ${\nu_{\rm min1}}$. 
${N_{\rm f}}$ takes a value of either two or three,  
${\sigma_{\rm i,obs}}$ are the observed uncertainty in $\braket{\Delta\nu}$,  ${\nu_{\rm min0}}$ and ${\nu_{\rm min1}}$,  ${P_{\rm i,obs}}$ are the observed non-asteroseismic data (${T_{\rm es}}$, $\log g$, and $R$), ${N_{\rm s}}$ is total number of included data points, and ${P_{\rm i,mod}}$ are the non-asteroseismic parameters of the model. 


 Uncertainties in $M$, $R$, $\log g$ and $t$  
reported in this study are estimated using Monte-Carlo simulations.
For each star, 
the method is repeated on each synthetic data set and the standard deviation of the distribution after 1000 iterations
is taken as an estimate of the uncertainty. 
The typical relative uncertainties 
obtained using this method  2-3, 3-4 percent  for $R$ and $M$, respectively.
As the uncertainty in $L$ is not computed by the Monte-Carlo simulations, we obtain them using a quadratic approach.
The uncertainty in ${L_{\rm mod}}$ is given by
\begin{equation}
\frac{\Delta L_{\rm mod}}{L_{\rm mod}}=\sqrt{\left(2\frac{\Delta R_{\rm mod}}{R_{\rm mod}}\right)^2+\left(4\frac{\Delta T_{\rm mod}}{T_{\rm mod}}\right)^2}.
\end{equation}
The parameter uncertainties are listed in Table 2. 
The uncertainty in ${T_{\rm mod}}$ is assumed to be equal to ${\Delta T_{\rm es}}$. 
Using model data, we obtained  ${\Delta Y_{\rm mod}}$/ ${Y_{\rm mod}}$$\approx$$3{\Delta M_{\rm mod}}$/${M_{\rm mod}}$; ${\Delta Z_{\rm mod}}$ is obtained in a similar manner.  

\begin{table*}
        \caption{Fundamental {\small MESA} model parameters of the evolved stars. $M_{\rm mod}$, $R_{\rm mod}$, ${T_{\rm mod}}$, $L_{\rm mod}$, ${\log g_{\rm mod}}$, $Y_{\rm mod}$, $Z_{\rm mod}$, $\alpha$, $t_{\rm mod}$, and ${{\chi^2_{\rm spec}}}$ are ordered between the second and eleventh columns. $M_{\rm mod}$, $R_{\rm mod}$ and $L_{\rm mod}$ are in solar units. While ${T_{\rm mod}}$ is in K, ${\log g_{\rm mod}}$ and $t_{\rm mod}$ are given in units of cgs and Gyr, respectively.
}
\small\addtolength{\tabcolsep}{-2pt}
        \begin{tabular}{lcccccccccr}
                \hline
         Star   & $M_{\rm mod}$  & $R_{\rm mod}$ & $T_{\rm mod}$ & $L_{\rm mod}$ & ${\log g_{\rm mod}}$& $Y_{\rm mod}$ &$Z_{\rm mod}$&  $\alpha$& $t_{\rm mod}$ & ${{\chi^2_{\rm spec}}}$ \\
              & $(M_{\sun})$ &$(R_{\sun})$ & (K)  & $(L_{\sun})$ &(cgs)& & & &(Gyr)&     \\
                \hline
HD~2151      &1.09$\pm$0.03&1.83$\pm$0.02&5872$\pm$45&3.58$\pm$0.13&3.95$\pm$0.02&0.298$\pm$0.024&0.0145$\pm$0.0012&2.00&6.33$\pm$0.75&0.13\\[2pt]
KIC~5955122  &1.17$\pm$0.04&2.09$\pm$0.02&5865$\pm$70&4.65$\pm$0.18&3.96$\pm$0.02&0.295$\pm$0.030&0.0103$\pm$0.0011&2.00&5.07$\pm$0.72&0.21\\[2pt]
KIC~7341231  &0.79$\pm$0.02&2.59$\pm$0.03&5366$\pm$50&4.99$\pm$0.23&3.51$\pm$0.02&0.260$\pm$0.020&0.0010$\pm$0.0008&2.31&12.75$\pm$2.29&0.29\\[2pt]
KIC~7747078  &1.10$\pm$0.03&1.92$\pm$0.02&5910$\pm$70&4.06$\pm$0.16&3.91$\pm$0.02&0.275$\pm$0.022&0.0110$\pm$0.0009&2.15&6.22$\pm$0.74&0.17\\[2pt]
KIC~7976303  &1.06$\pm$0.03&1.97$\pm$0.02&6050$\pm$70&4.67$\pm$0.21&3.87$\pm$0.02&0.298$\pm$0.025&0.0085$\pm$0.0007&2.43&5.57$\pm$0.65&0.01\\[2pt]
KIC~8228742  &1.15$\pm$0.03&1.78$\pm$0.02&6042$\pm$70&3.79$\pm$0.17&3.99$\pm$0.02&0.298$\pm$0.023&0.0145$\pm$0.0011&1.76&4.73$\pm$0.55&0.01\\[2pt]
KIC~8524425  &1.07$\pm$0.03&1.79$\pm$0.02&5614$\pm$70&2.87$\pm$0.14&3.87$\pm$0.02&0.284$\pm$0.024&0.0175$\pm$0.0015&2.00&8.51$\pm$1.00&0.11\\[2pt]
KIC~8561221  &1.47$\pm$0.04&3.15$\pm$0.04&5243$\pm$70&6.74$\pm$0.35&3.61$\pm$0.02&0.298$\pm$0.024&0.0140$\pm$0.0011&1.85&2.19$\pm$0.27&0.02\\[2pt]
KIC~8702606  &1.27$\pm$0.04&2.49$\pm$0.03&5442$\pm$70&2.49$\pm$0.19&3.75$\pm$0.02&0.270$\pm$0.026&0.0135$\pm$0.0013&2.00&4.39$\pm$0.51&0.06\\[2pt]
KIC~10018963 &1.16$\pm$0.03&1.92$\pm$0.02&6216$\pm$112&4.96$\pm$0.37&3.93$\pm$0.02&0.298$\pm$0.023&0.0116$\pm$0.0009&2.00&4.18$\pm$0.48&0.27\\[2pt]
KIC~10920273 &1.06$\pm$0.03&1.82$\pm$0.02&5800$\pm$75&3.36$\pm$0.20&3.94$\pm$0.02&0.298$\pm$0.025&0.0140$\pm$0.0012&2.00&7.01$\pm$0.86&0.02\\[2pt]
KIC~11026764 &1.11$\pm$0.03&2.03$\pm$0.02&5688$\pm$70&3.88$\pm$0.27&3.87$\pm$0.02&0.288$\pm$0.023&0.0150$\pm$0.0015&2.10&6.19$\pm$0.74&0.01 \\[2pt]
KIC~11244118 &1.09$\pm$0.03&1.60$\pm$0.02&5731$\pm$70&2.47$\pm$0.13&4.07$\pm$0.02&0.269$\pm$0.022&0.0200$\pm$0.0017&2.10&9.12$\pm$1.12&0.25\\[2pt]
KIC~11395018 &1.13$\pm$0.03&2.11$\pm$0.02&5459$\pm$85&3.54$\pm$0.25&3.84$\pm$0.02&0.294$\pm$0.023&0.0170$\pm$0.0014&2.00&6.46$\pm$0.80&0.11\\[2pt]
KIC~11414712 &1.06$\pm$0.03&2.19$\pm$0.03&5629$\pm$70&4.34$\pm$0.23&3.78$\pm$0.02&0.283$\pm$0.024&0.0085$\pm$0.0007&1.80&6.18$\pm$0.72&0.23\\[2pt]

               \hline
        \end{tabular}
\end{table*}

\section{Results and Discussions}
The best-fitting estimates of the actual observables are listed in Table 2.
Because models $l=1$ and $2$ are mixed modes and therefore have no available reference frequencies, we limited our analysis to the $l=0$ modes of the observed 
oscillation frequencies. 
We
also computed the fundamental parameters of each stars using conventional and modified scaling relations. 

\subsection{$\Delta\nu$ of evolved stars}
 The large separation, $\Delta\nu$, is the frequency difference between modes with the same degree of consecutive orders. The $\Delta\nu$ of a star is approximately proportional to the square root of its mean density (Ulrich 1986). The mean $\Delta\nu$ ($\braket{\Delta\nu_{\rm obs}}$) values obtained from the observed and model oscillation frequencies ($\braket{\Delta\nu_{\rm mod}}$) of the stars are listed in Tables 1 and 3, respectively. 
\begin{figure}
\begin{center}
\includegraphics[width=\columnwidth]{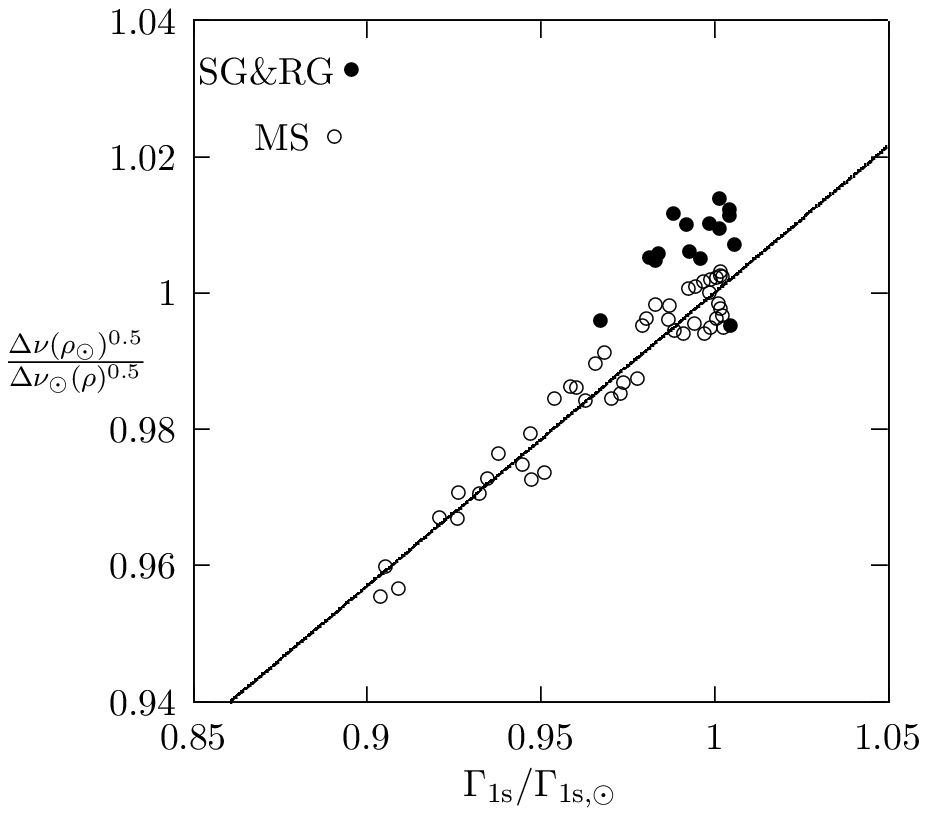}
\caption{Plot of ($\braket{\Delta\nu}$/$\braket{\Delta\nu_{\sun}}$)/($\rho$/$\rho_{\sun}$)$^{0.5}$ as function of $\Gamma_{\rm 1s}$/$\Gamma_{\rm 1s\sun}$ for evolved stars and MS models. The circles and filled circles represented data from the MS models in Paper III and evolved stars, respectively. The fitting line for the MS models is 0.430 $\Gamma_{\rm 1s}$/$\Gamma_{\rm 1s\sun}$+0.570.}

\end{center}
\end{figure}

\par In Paper III, it is shown that the first adiabatic exponent at the surface of a star (${\Gamma_{\rm 1s}}$) affects the relation between $\braket{\Delta\nu}$ and the square root of its mean density ($\braket{\rho}$). 
 ${\Gamma_{\rm 1s}}$ is influenced by the ionization state of the gas.  
The ratio of $\braket{\Delta\nu}$ to $\braket{\rho}$ in solar units is defined as $f_{\rm \Delta\nu}$; according to the MS models with various masses, $f_{\rm \Delta\nu}$ is a linear function of 
${\Gamma_{\rm 1s}}$, namely, $\Gamma_{\rm 1s}$/$\Gamma_{\rm 1s\sun}$ :
\begin{equation}
f_{\Delta \nu}=\frac{\braket{\Dnu}/\braket{\Dnu_{\sun}}}{(\braket{\rho}/\braket{\rho_{\sun}})^{1/2}}=0.430\frac{\Gamma_{\rm \negthinspace 1s}}{\Gamma_{\rm \negthinspace 1s\odot}}+0.570,
\end{equation}
where $\braket{\Dnu_{\sun}}$=136 $\mu$Hz and $\Gamma_{\rm \negthinspace 1s\odot}$=1.639 are the solar values (Paper III).
 
In Paper III, $\braket{\Dnu_{\sun}}$ is taken as 136 ${\rm \mu}$Hz in derivation of these relations. In Eq. 2 we define general scaling relation, so we prefer to take $\braket{\Dnu_{\sun}}$= 135.15 ${\rm \mu}$Hz 
(Broomhall et al. 2009).

The values of $f_{\Delta \nu}$ for the best-fitting models are plotted as function of ${\Gamma_{\rm \negthinspace 1s}}$ in Fig. 2, in which the fitting formula given in equation (6) is represented by the solid line.
The evolved stars and the MS models from Paper III are represented by filled circles and circles, respectively. 
All of SGs and the RG (KIC 8561221) are located slightly above the MS models and clustered around a 
${\Gamma_{\rm \negthinspace 1s}}/{\Gamma_{\rm \negthinspace 1s\odot}}$  value of 1. 


\begin{figure}
\begin{center}
\includegraphics[width=\columnwidth]{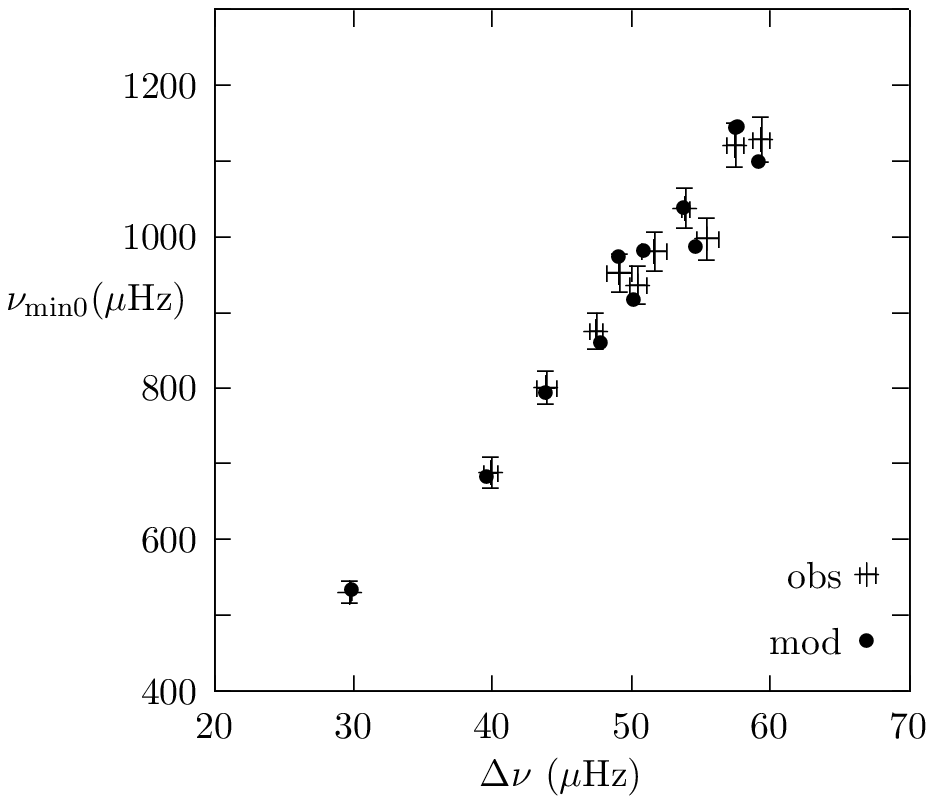}
\caption{ $\nu_{\rm min0}$ of the evolved stars is plotted with respect to $\braket{\Delta\nu}$.
The filled circles and pluses represent values of  $\nu_{\rm min0}$ obtained from model and observational oscillation frequencies, respectively.}
\end{center}
\end{figure}

\subsection{$\nu_{\rm min}$ of evolved stars}
${\Delta\nu}$ of the model oscillation frequencies and observed frequencies of each star are plotted as a function of $\nu$ (see Fig. 9).
Based on this comparison, we are able to ensure that the observed and model minimum frequencies are close to each other. In addition to fitting the observed ${\Delta\nu}$ values to the models, more precise models could be obtained by fitting them to the observed minima values.

The ${{\chi^2_{\rm seis}}}$ values obtained form  equation (4) are used to assess the similarities between the model and the observed oscillation frequencies interms of $\nu_{\rm min}$s
and the $\Delta\nu$. 
The results obtained in this process are listed in Table 3.
\begin{table*}
        \caption{Asteroseismic parameters of {\small {MESA}} model results for the evolved stars.
                 ${\braket{{\delta{\nu_{02,\rm mod}}}}}$, ${\braket{\Delta{\nu_{\rm mod}}}}$, ${\nu_{\rm max,mod}}$,
                 ${\nu_{\rm min0,mod}}$, ${\nu_{\rm min1,mod}}$ and ${\nu_{\rm min2,mod}}$ indicate, respectively, the mean small and large separations between the model oscillation frequencies,
                 the model oscillation at the frequency of the maximum amplitude, and the reference frequencies of models in $\mu$Hz units. ${\nu_{\rm max,mod}}$ are computed from the scaling relations with
                 ${T_{\rm mod}}$ and ${logg_{\rm mod}}$. The model ${{\chi^2_{\rm seis}}}$ values are given in the final column.Typical uncertainties for the reference frequencies are ${\braket{\Delta{\nu_{\rm mod}}}}/2$. }
        \begin{tabular}{lcrrrrrr}
                \hline
         Star   &  ${\braket{{\delta{\nu_{02,\rm mod}}}}}$ & ${\braket{\Delta{\nu_{\rm mod}}}}$ & ${\nu_{\rm max,mod}}$ & ${\nu_{\rm min0,mod}}$ &
          ${\nu_{\rm min1,mod}}$ & ${\nu_{\rm min2,mod}}$& ${{\chi^2_{\rm seis}}}$ \\
         &  ($\mu$Hz) & ($\mu$Hz) & ($\mu$Hz) & ($\mu$Hz) & ($\mu$Hz) & ($\mu$Hz) &  \\
                \hline
HD~2151      &4.7&57.7&984.8&1144.9&883.0&652.9&0.79\\[2pt]
KIC~5955122  &4.7&49.1&810.8&973.5&720.1&553.2&0.72\\[2pt]
KIC~7341231  &---&29.0&434.9&---&358.6&---&0.68\\[2pt]
KIC~7747078  &3.9&53.8&896.9&1037.3&791.8&567.0&0.26\\[2pt]
KIC~7976303  &4.5&50.9&814.0&980.4&770.3&548.1&0.18\\[2pt]
KIC~8228742  &4.4&61.7&1082.5&1483.4&1045.6&744.0&0.41\\[2pt]
KIC~8524425  &5.1&59.2&1059.4&1098.3&881.5&612.0&0.29\\[2pt]
KIC~8561221  &2.6&29.7&474.3&533.4&397.9&308.8&0.32\\[2pt]
KIC~8702606  &3.5&39.6&646.3&682.3&537.1&410.1&0.21\\[2pt]
KIC~10018963 &4.7&54.7&922.8&986.7&781.2&---&0.14\\[2pt]
KIC~10920273 &5.3&57.5&978.4&1142.5&880.5&652.9&0.16\\[2pt]
KIC~11026764 &4.7&50.2&831.0&915.8&715.6&518.5&0.73\\[2pt]
KIC~11244118 &5.6&71.0&1306.8&---&1153.9&849.3&0.15\\[2pt]
KIC~11395018 &4.2&47.8&798.6&859.4&662.4&495.1&0.41\\[2pt]
KIC~11414712 &4.1&43.8&682.9&793.3&613.4&449.1&0.61\\[2pt]
                \hline
        \end{tabular}
\end{table*}

In Fig. 3, the observed and model minimum frequencies, $\nu_{\rm min0,obs}$ (filled circles) and $\nu_{\rm min0,mod}$ (pluses) are compared. 
For KIC 7341231 and KIC 11244118, the values
of {$\nu_{\rm min0}$ corresponding to the high frequencies could be
obtained from neither the observations nor the models. There are good agreement between model and observational values of $\nu_{\rm min0}$. The same is true for  $\nu_{\rm min1}$.

Fig. 4 compares the values of $\nu_{\rm min1,obs}$ (filled circles) and $\nu_{\rm min1,mod}$ (pluses). 
It is seen from the figure that the observed and model oscillation frequencies are in excellent agreement, with the model values generally falling within the error ranges of the observed $\nu_{\rm min1}$ values.

\begin{figure}
\begin{center}
\includegraphics[width=\columnwidth]{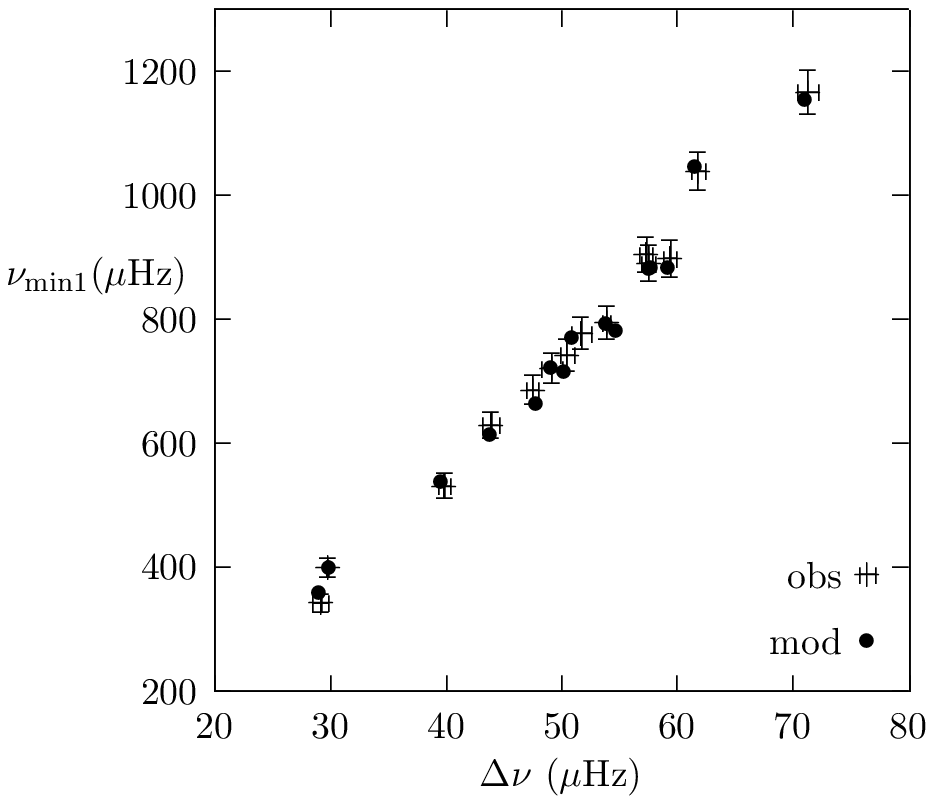}
\caption{$\nu_{\rm min1}$ of the evolved stars is plotted with respect to $\braket{\Delta\nu}$.
The filled circles and pluses represent values of  $\nu_{\rm min1}$ obtained from model and observational oscillation frequencies, respectively.}
\end{center}
\end{figure}


\subsection{Comparison of radii}

\begin{figure}
\begin{center}
\includegraphics[width=\columnwidth]{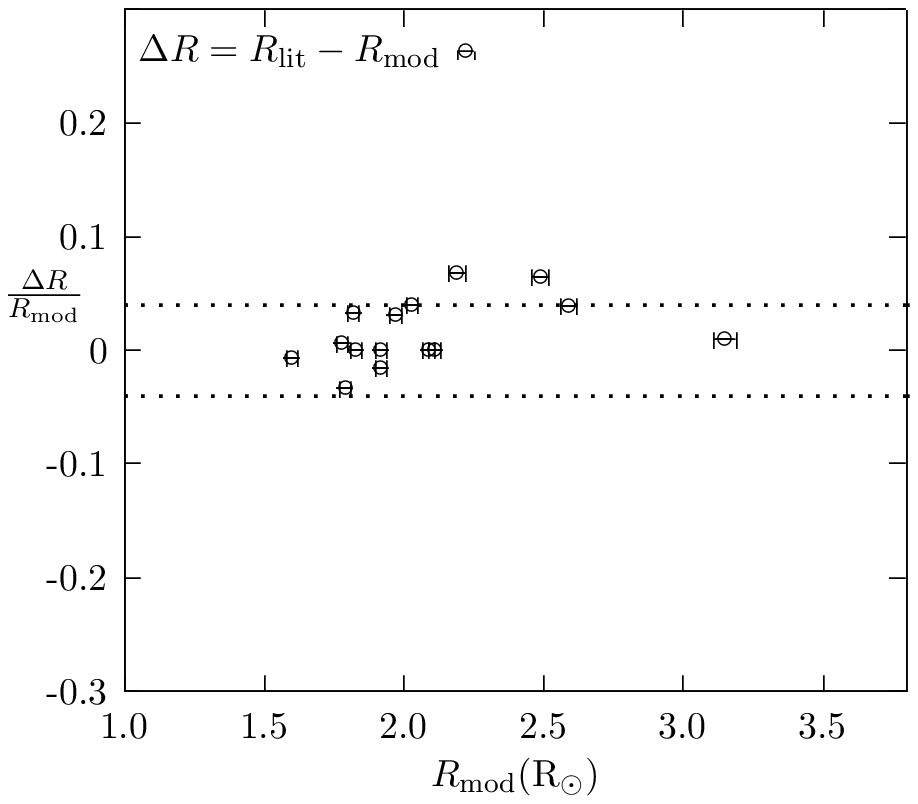}
   \caption{ $R_{\rm lit}-R_{\rm mod}$ is plotted with respect to $R_{\rm mod}$ in solar units.
The dotted lines are for -0.04 and 0.04.
}
\end{center}
\end{figure}

\begin{figure}
\begin{center}
\includegraphics[width=\columnwidth]{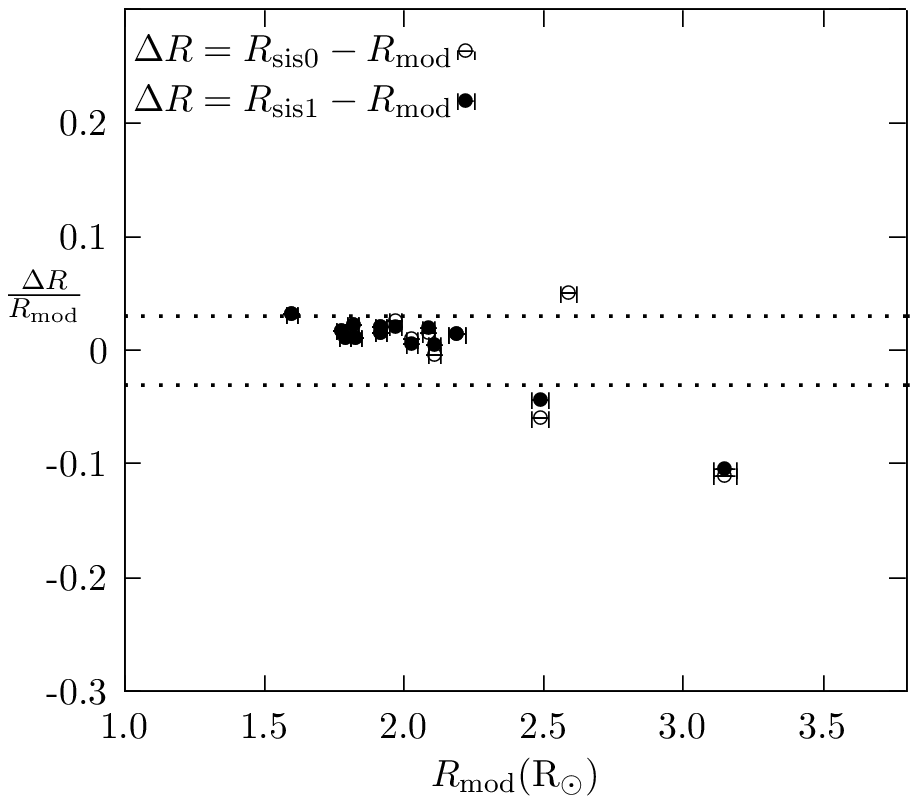}
   \caption{$(R_{\rm sis0}-R_{\rm mod})/R_{\rm mod}$ (circles) and $(R_{\rm sis1}-R_{\rm mod})/R_{\rm mod}$ (filled circles) are plotted with respect to $R_{\rm mod}$ in solar units.
The dotted lines are for -0.03 and 0.03.
}
\end{center}
\end{figure}

\par The oscillation frequencies observed in a star depend on the its mean density and the mean density is inversely proportional to the radius.
By fitting the observed oscillation frequencies with those obtained from the model, it is possible to determine the stellar radius very precisely from the interior models.
Fig. 5, shows a comparison of the stellar radii produced by  {\small {MESA}} model ($R_{\rm mod}$),
with those obtained  from 
the literature ($R_{\rm lit}$).
For majority of stars, they are in good agreement, so that the difference is less than 4 per cent. 
Significant discrepancy appears for KIC 8702606 and KIC 11414712. For these two stars, the difference between $R_{\rm lit}$ and $R_{\rm mod}$ is about 6  per cent.

In Paper IV, fitting formulae are derived for radius as a function of $\nu_{\rm min0}$ ($R_{\rm sis0}$) and $\nu_{\rm min1}$ ($R_{\rm sis1}$) from the MS models.
 In Fig. 6, $(R_{\rm sis0}-R_{\rm mod})/R_{\rm mod}$ and $(R_{\rm sis1}-R_{\rm mod})/R_{\rm mod}$ are plotted with respect to $R_{\rm mod}$. Both of $R_{\rm sis0}$ and $R_{\rm sis1}$ (MS models) are in very good agreement with $R_{\rm mod}$ (post-MS models) 
if $R_{\rm mod}< 2.2 R_{\sun}$, despite the different evolutionary phases.
For these stars $R_{\rm sis0}$ and $R_{\rm sis1}$ are slightly greater than $R_{\rm mod}$, about 1.5 per cent ($\approx 0.029 R_{\sun}$).
This implies that the derived formulae for $R_{\rm sis0}$ and $R_{\rm sis1}$ are also suitable for such stars, namely SGs. 
For the larger stars, the difference becomes larger. The three stars with $R_{\rm mod}> 2.2 R_{\sun}$ (KIC 7341231, KIC 8561221 and KIC 8561221) have $T_{\rm eff} < 5450$ K and $\log g < 3.77$.

The model results and the literature values for mass and radius are summarized in Table 4.
For comparison, masses and radii found in Papers III and IV are also listed.

 

\subsection{Comparison of ${T_{\rm eff}}$s}
\begin{figure}
\begin{center}
\includegraphics[width=\columnwidth]{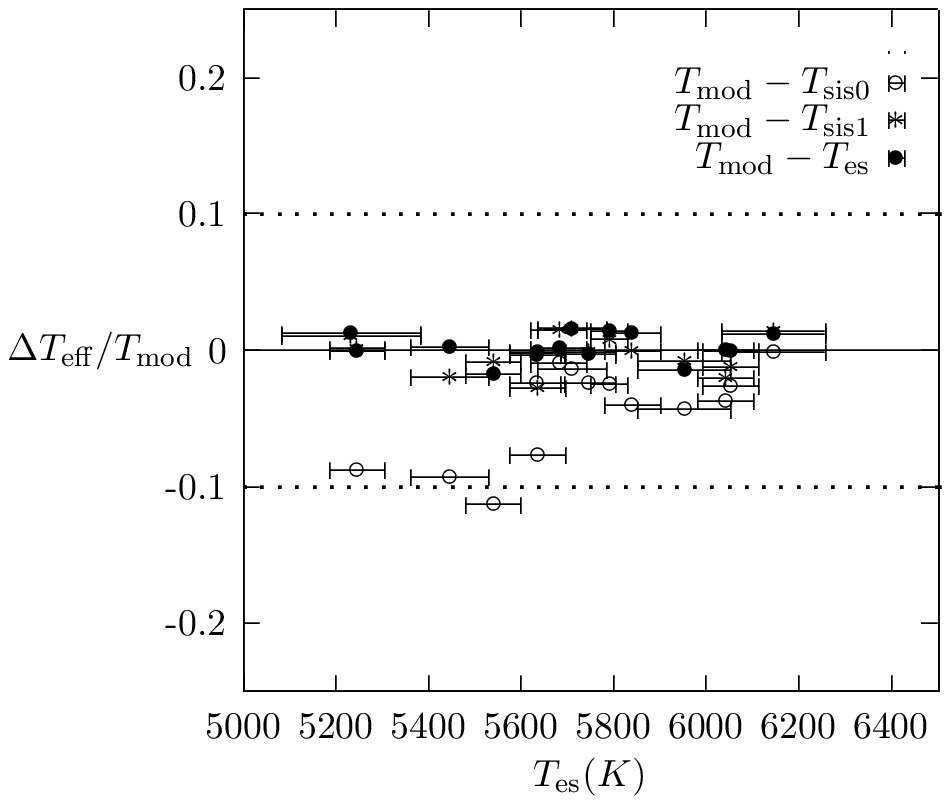}
   \caption{Comparison of ${T_{\rm mod}}$ obtained from calibration of interior models with ${T_{\rm es}}$ and asteroseismically derived ${T_{\rm eff}}$ values from Paper IV. 
The circles and asterisks indicate $T_{\rm mod}-{T_{\rm sis0}}$ and $T_{\rm mod}-{T_{\rm sis1}}$, respectively.
${T_{\rm min0}}$ and ${T_{\rm min1}}$ are computed from the fitting formula based on $\nu_{\rm min0}$ and $\nu_{\rm min1}$, respectively. 
$T_{\rm mod}-{T_{\rm es}}$ is represented by the filled circles. 
}
\end{center}
\end{figure}
\begin{table*} 
        \caption{Listing of masses and radii of all the evolved stars in this study. $M_{\rm sca}$ and $R_{\rm sca}$ are determined from conventional scaling relations. $M_{\rm III}$ and $R_{\rm III}$, and $M_{\rm IV}$ and $R_{\rm IV}$  are calculated using formulas from Paper III and IV, respectively. 
 {\small {MESA}} masses and radii are given as $M_{\rm mod}$ and $R_{\rm mod}$, respectively. $M_{\rm lit}$ and $R_{\rm lit}$ are masses and radii, respectively, from the literature.  
                }
\small\addtolength{\tabcolsep}{-2pt}
        \begin{tabular}{lcrrrrrrrrrrrrrr}
                \hline
         Star   & $M_{\rm sca}$ & $R_{\rm sca}$ & $M_{\rm III}$ & $R_{\rm III}$&  $M_{\rm IV}$ & $R_{\rm IV}$& 
           $M_{\rm mod}$ & $R_{\rm mod}$ &
           $M_{\rm lit}$ & $R_{\rm lit}$
\\
         & $M_{\sun}$ &$R_{\sun}$ &$M_{\sun}$ &$R_{\sun}$&$M_{\sun}$ &$R_{\sun}$ &$M_{\sun}$ &$R_{\sun}$ &$M_{\sun}$ &$R_{\sun}$
\\
                \hline
HD~2151      &1.01$\pm$0.01&1.75$\pm$0.02&1.01$\pm$0.15&1.75$\pm$0.10&1.10$\pm$0.15&1.83$\pm$0.03&1.09$\pm$0.03&1.83$\pm$0.02&1.09$\pm$0.05&1.83$\pm$0.03\\[1.2pt]
KIC~5955122  &1.18$\pm$0.23&2.06$\pm$0.15&1.16$\pm$0.24&2.05$\pm$0.16&1.25$\pm$0.03&2.11$\pm$0.02&1.17$\pm$0.04&2.09$\pm$0.02&1.12$\pm$0.02&2.04$\pm$0.03\\[1.2pt]
KIC~7341231  &0.7$\pm$0.159&2.58$\pm$0.20&0.79$\pm$0.10&2.58$\pm$0.21&0.81$\pm$0.04&2.69$\pm$0.20&0.79$\pm$0.02&2.59$\pm$0.03&0.90$\pm$0.10&2.69$\pm$0.20\\[1.2pt]
KIC~7747078  &1.26$\pm$0.17&1.99$\pm$0.10&1.26$\pm$0.17&1.99$\pm$0.10&1.12$\pm$0.02&1.94$\pm$0.10&1.10$\pm$0.03&1.92$\pm$0.02&1.06$\pm$0.05&1.89$\pm$0.02\\[1.2pt]
KIC~7976303  &1.45$\pm$0.15&2.17$\pm$0.10&1.42$\pm$0.15&2.14$\pm$0.10&1.15$\pm$0.02&2.11$\pm$0.01&1.06$\pm$0.03&1.97$\pm$0.02&1.17$\pm$0.02&2.03$\pm$0.01\\[1.2pt]
KIC~8228742  &1.42$\pm$0.18&1.90$\pm$0.10&1.39$\pm$0.19&1.87$\pm$0.10&1.24$\pm$0.02&1.81$\pm$0.01&1.15$\pm$0.03&1.78$\pm$0.02&1.31$\pm$0.01&1.79$\pm$0.01\\[1.2pt]
KIC~8524425  &1.11$\pm$0.15&1.78$\pm$0.09&1.12$\pm$0.16&1.79$\pm$0.09&1.10$\pm$0.02&1.79$\pm$0.01&1.07$\pm$0.03&1.79$\pm$0.02&1.00$\pm$0.07&1.73$\pm$0.02\\[1.2pt]
KIC~8561221  &1.39$\pm$0.10&3.07$\pm$0.08&1.38$\pm$0.10&3.07$\pm$0.08&1.56$\pm$0.01&3.19$\pm$0.01&1.47$\pm$0.04&3.15$\pm$0.04&1.55$\pm$0.13&3.18$\pm$0.18\\[1.2pt]
KIC~8702606  &1.31$\pm$0.19&2.47$\pm$0.14&1.32$\pm$0.19&2.48$\pm$0.14&1.13$\pm$0.02&2.37$\pm$0.01&1.27$\pm$0.04&2.49$\pm$0.03&1.27$\pm$0.03&2.65$\pm$0.02\\[1.2pt]
KIC~10018963 &1.34$\pm$0.20&1.99$\pm$0.12&1.30$\pm$0.21&1.96$\pm$0.12&1.24$\pm$0.02&1.95$\pm$0.01&1.16$\pm$0.03&1.92$\pm$0.02&1.18$\pm$0.03&1.92$\pm$0.02\\[1.2pt]
KIC~10920273 &1.07$\pm$0.20&1.81$\pm$0.17&1.08$\pm$0.30&1.82$\pm$0.17&1.11$\pm$0.02&2.03$\pm$0.01&1.06$\pm$0.03&1.82$\pm$0.02&1.00$\pm$0.04&1.88$\pm$0.02\\[1.2pt]
KIC~11026764 &1.28$\pm$0.21&2.10$\pm$0.13&1.29$\pm$0.21&2.11$\pm$0.13&1.13$\pm$0.02&2.03$\pm$0.01&1.11$\pm$0.03&2.03$\pm$0.02&1.27$\pm$0.06&2.11$\pm$0.03\\[1.2pt]
KIC~11244118 &1.07$\pm$0.18&1.57$\pm$0.09&1.08$\pm$0.18&1.57$\pm$0.09&1.21$\pm$0.02&1.64$\pm$0.01&1.09$\pm$0.03&1.60$\pm$0.02&1.10$\pm$0.05&1.59$\pm$0.03\\[1.2pt]
KIC~11395018 &1.27$\pm$0.31&2.17$\pm$0.19&1.28$\pm$0.31&2.17$\pm$0.19&1.11$\pm$0.02&2.10$\pm$0.01&1.13$\pm$0.03&2.11$\pm$0.02&1.13$\pm$0.04&2.11$\pm$0.02\\[1.5pt]
KIC~11414712 &1.15$\pm$0.18&2.23$\pm$0.14&1.16$\pm$0.18&2.24$\pm$0.14&1.08$\pm$0.03&2.18$\pm$0.02&1.06$\pm$0.03&2.19$\pm$0.03&1.26$\pm$0.08&2.34$\pm$0.06\\[1.2pt]
                \hline
        \end{tabular}
\end{table*}

\par ${T_{\rm eff}}$ is another important parameter in asteroseismic scaling relations for stellar mass and radius. 
${T_{\rm es}}$, obtained from stellar spectrum, has an uncertainty of nearly 200 K (Bruntt et al. 2012), which has a critical role in the scaling relations (see equations 1 and 2). For example, if the effective temperature of a star is ${T_{\rm es}}=5800$ K with an uncertainty of 100 K, the uncertainty in the mass and radius values obtained from the scaling relations will be 2.6 and 0.9 per cent, respectively.

\par In Paper IV, the effective temperatures of 90 solar-like oscillating stars are computed using five different methods: the obtained results are ${T_{\rm eVK}}$ (from V-K colour), ${T_{\rm eBV}}$ (from B-V color), ${T_{\rm sis0}}$ (from ${\nu_{\rm min0}}$), ${T_{\rm sis1}}$ (from ${\nu_{\rm min1}}$) and ${T_{\rm sis2}}$ (from ${\nu_{\rm min2}}$). 
Fig. 7 shows a comparison of ${T_{\rm mod}}$ with ${T_{\rm es}}$, ${T_{\rm sis0}}$ and ${T_{\rm sis1}}$.  
Overall, the ${T_{\rm sis1}}$ and ${T_{\rm es}}$ values are in excellent agreement with the corresponding ${T_{\rm mod}}$ values, with a maximum difference of less than approximately one percent.
For ${T_{\rm sis0}}$, however there are significant differences between ${T_{\rm sis0}}$ and ${T_{\rm mod}}$. The reason behind this discrepancy should be investigated.


The degree of agreement with the non-asteroseismic observational parameters and model parameters is measured by calculating ${{\chi^2_{\rm spec}}}$ using equation (3).
In this process, the ${T_{\rm eff}}$ and $\log g$ determined from the spectrum are taken into consideration.

\subsection{Comparison of ages}
\begin{figure}
\begin{center}
\includegraphics[width=\columnwidth]{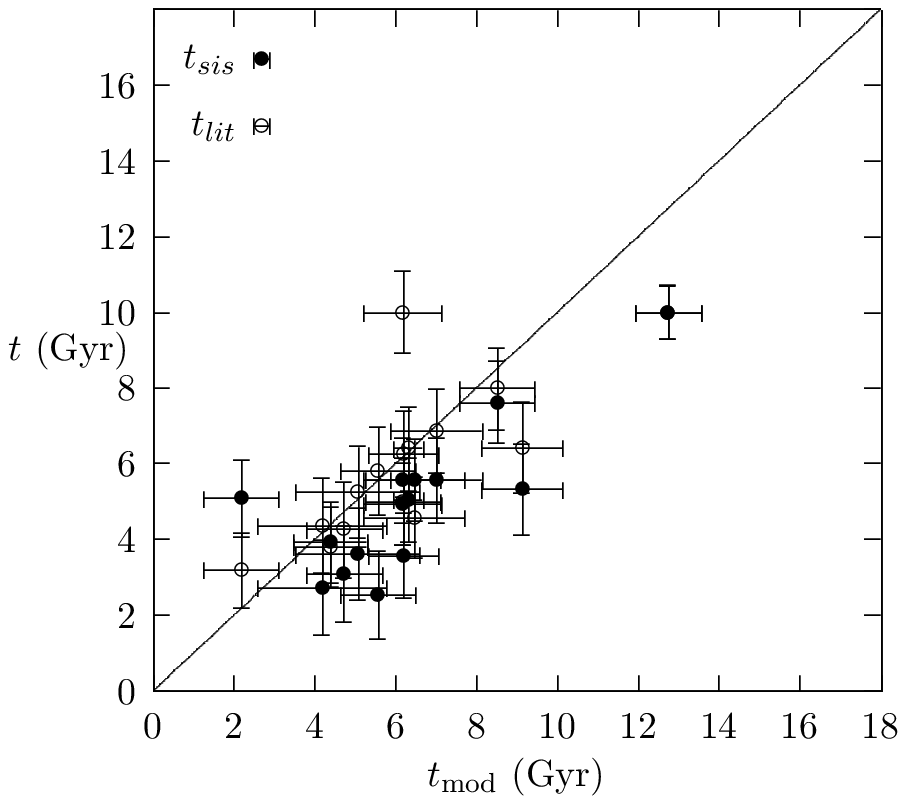}
   \caption{Comparison of stellar ages obtained using the {\small {MESA}} models and the ages from the literature. 
Uncertainty of $t_{\rm lit}$ is not available for KIC 8702606, KIC 7341231 and KIC 11414712.
Filled circles are the ages from Li et al. (2020).
} 
\end{center}
\end{figure}
 Stellar age is one of the essential parameters for the understanding of stellar structure. The age of star cannot be directly determined using observational data but must instead be estimated using an interior model. 
As many different models 
for reconstructing age from limited observational data are available, a given star will have conflicting age estimates. To delimit age of stars  
in our study, we used observed asteroseismic and non-asteroseismic data.

Fig. 8 shows the estimated ages of the target stars obtained from interior models ($t_{\rm mod}$) versus ages in the literature ($t_{\rm lit}$). 
$t_{\rm lit}$ is taken from Deheuvels et al. (2014), Metcalfe et al. (2014), Garcia et al. (2014), Do\u{g}an et al. (2013) and Brandao et al. (2011). 
For majority of the stars, $t_{\rm lit}$ and $t_{\rm mod}$ are in good agreement, the difference between them is less than 1 Gyr for the nine targets.
Also shown in Fig. 8 is the recent ages ($t_{\rm Li}$) obtained by Li et al. (2020) for ten stars. The greatest difference between $t_{\rm Li}$ and $t_{\rm mod}$ is for KIC 10920273,
about 5.19 Gyr.
However, $t_{\rm mod}-t_{\rm lit}$ is just 0.15 Gyr for this star.

The stellar mass has the most significant influence on the determined age of a star. 
Although the masses of stars calculated in 
the literature differ from those calculated using {\small {MESA}}, the differences between the two ages are generally only a few Gyr. 
Based on this, the model oscillation frequencies could be fitted to the observed oscillation frequencies to obtain age estimates that are independent of the input parameters of the model. 

\section{Conclusions}
In this study, we constructed models of 15 evolved stars using the {\small {MESA}} code
and used the {\small {MESA}} models in conjunction with the stars' observed oscillation frequencies to determine the fundamental properties of the stars.
As oscillation frequency data, we used observations from both a ground-based telescope and the \textit{Kepler} space telescope.
The observed asteroseismic ($\Delta\nu$, ${\nu_{\rm max}}$,  ${\nu_{\rm min1}}$,  ${\nu_{\rm min2}}$)  and non-asteroseismic ($T_{\rm eff}$, $\log g$, [Fe/H])  parameters are compared with model properties 
obtained using the {\small {MESA}} code.
The 15
evolved stars are estimated to have masses and radii within the ranges of
$0.79-1.47$ $M_{\rm \sun}$ and $1.60-3.15$ $R_{\rm \sun}$, respectively, with typical uncertainties of
$\sim$ 3-6 per cent and $\sim$ 1-2 per cent in terms of mass and radius, respectively.
 By fitting the reference frequencies, typically the accuracy of the asteroseismic radius, mass, and age is much better than that determined from $\Dnu$ and $\nu _{\rm max}$.
Also, the stellar $R$ and $M$ determined from the interior models are more accurate than the $R$ and $M$ obtained from the different scaling relations.
 
Metallicity is crucial to the understanding of the evolution and structure of stars. 
In the literature, the metallicity of a star is generally calculated from its [Fe/H]
value, a factor that might not in fact represent the exact metallicity of the star. In this study, metallicity is calculated from the relation between [O/H] and [Fe/H] obtained from Y{\i}ld{\i}z et al. (2014b).
Based on this metallicity, the stellar interior models are constructed using the {\small {MESA}} evolution code.

It is quite difficult to determine the age of a star from the observations;
instead, the age of individual stars must be calculated from interior models.
The results of these models should be tested 
using different methods or observed data. 
The estimated ages for the 15 evolved stars in this study (Table 2) are all compared with ages available in the literature. In addition, the values of
$M$, Z, mixing-length parameter ($\alpha$), and $Y$
determined using {\small {MESA}}
are found to vary significantly from those produced by models in the literature. However, values of $t_{\rm lit}$
(Fig. 8) are found to be in close agreement with $t_{\rm mod}$.
The difference between the ages obtained from 
different evolution codes provided correct ages within a range of
several Gyr.

Determining the asteroseismic scaling relations for solar-like stars
is very important because
they can be used to derive 
the fundamental parameters of the single stars from 
observed properties ($\braket{\Delta\nu}$, ${\nu_{\rm max}}$ and $T_{\rm eff}$). In this study, 
we tested the scaling relations for the selected evolved stars. 
In doing so, we attempted to fit the reference frequencies,
$\braket{\Delta\nu}$, and ${\nu_{\rm max}}$
to ensure compatibility between the model and observed oscillation frequencies.
In particular, the models that are constructed to fit the minima provided much more accurate results.

It is found that the asteroseismic effective temperature 
${T_{\rm sis1}}$ 
are in good agreement with ${T_{\rm es}}$ and 
${T_{\rm mod}}$. However, there is significant temperature difference between ${T_{\rm sis0}}$ and ${T_{\rm mod}}$ (or ${T_{\rm es}}$).

Despite the different evolutionary phases, new expressions obtained for $R_{\rm sis0}$ and $R_{\rm sis1}$ from MS models in Paper IV give very consistent results with $R_{\rm mod}$, 
if $R_{\rm mod}< 2.2 R_{\sun}$. 
For these stars $R_{\rm sis0}$ and $R_{\rm sis1}$ are only 1.5 per cent greater than $R_{\rm mod}$. Much more explicit formulae can be developed for $R_{\rm sis0}$ and $R_{\rm sis1}$ for SGs and RGs.

\section*{Acknowledgments}
We would like to thank  
Kelly Spencer for her kind help in checking the language of the
revised manuscript.
We are grateful to Ege University Planning and Monitoring Coordination of Organizational Development and Directorate of Library and Documantation for their support in editing and proofreading service of this study.
I thank also my son Emirhan Sami for his support.
~\\
~\\
{\uppercase{\bf{Data availability}}} \\
~\\
{The data underlying this article will be shared on reasonable request to the corresponding author.
} \\











\bsp	
\label{lastpage}
\end{document}